\documentclass[aps,prx,twocolumn,superscriptaddress,longbibliography,nobalancelastpage,floatfix, 10pt]{revtex4-2}
\usepackage{graphicx} 
\usepackage[hypertexnames=true]{hyperref}
\usepackage{siunitx}
\usepackage{physics}
\usepackage{orcidlink}
\usepackage{lipsum}
\usepackage{float}
\usepackage{dsfont}
\usepackage{amsmath}
\begin{document}

\title{Efficient construction of time‑invariant process tensors for simulating high‑dimensional non-Markovian open quantum systems}

\author{{\'E}mile Cochin}
\affiliation{Département de Physique de l'ENS Lyon, ENS de Lyon, 69007 Lyon, France}
\affiliation{SUPA, School of Physics and Astronomy, University of St Andrews, St Andrews, KY16 9SS, United Kingdom}
\affiliation{Sorbonne Université, Institut des NanoSciences de Paris, 4 place Jussieu, 75005 Paris, France}
\author{Jonathan Keeling}
\affiliation{SUPA, School of Physics and Astronomy, University of St Andrews, St Andrews, KY16 9SS, United Kingdom}
\author{Brendon W. Lovett}
\affiliation{SUPA, School of Physics and Astronomy, University of St Andrews, St Andrews, KY16 9SS, United Kingdom}
\author{Alex W. Chin}
\affiliation{Sorbonne Université, Institut des NanoSciences de Paris, 4 place Jussieu, 75005 Paris, France}
\affiliation{CNRS, Institut des NanoSciences de Paris, 4 place Jussieu, 75005 Paris, France}
\date{\today}
\begin{abstract}
    Numerical methods for obtaining exact dynamics of non-Markovian open quantum systems are mostly limited to either small systems or to short-time evolution only. Here, we propose a new algorithm for computing process tensors---matrix product operator (MPO) representations that capture the environment influence---which achieves greatly enhanced computational scalings with system size, while maintaining linear scaling with simulation length. We build on recent developments in the field which allow for long-time evolutions through process tensors which have a time-translational invariance. These can be built for general Gaussian environments and generic coupling operators with the system using infinite time-evolving block decimation (iTEBD). We introduce a modified iTEBD algorithm using intermediate compression steps which bring down the computation time scaling with system size $d$ from $\order{d^8}$ to $\order{d^4}$, as well as significantly lowering the required memory. To illustrate the power of this method, we apply it to the problem of dispersive qubit readout in circuit QED, which was previously out-of-reach numerically. The full treatment of the measurement resonator, which requires a large system space, combined with the long simulation times precipitated by the separation of timescales between the measurement drive and the environment dissipation, is now possible. The algorithm we introduce not only allows for capturing non-Markovian dynamics in large open quantum systems, but also further extends all the existing capabilities of process tensors, for example in quantum optimal control, or in computation of multi-time correlations or of steady states, to more complex systems with tens of levels.
\end{abstract}
\maketitle

\section{Introduction}
The framework of open quantum systems aims to describe the interaction of quantum systems with their environments, which leads experimental systems to undergo decoherence and dephasing~\cite{breuer_Theory_2007}. For the sake of simplicity, most descriptions of open quantum systems have been based on deriving quantum master equations for the system from so-called Born and Markov approximations. While these greatly reduce the theoretical and numerical complexities, a growing number of applications require an exact---non-Markovian---treatment of the environment in order to explain important features of  the system's dynamics~\cite{vagov_Real_2011,chin_role_2013a,barford_Electronic_2013,groblacher_Observation_2015a,potocnik_Studying_2018a}. 

The vast majority of studies of such non-Markovian open quantum systems  rely heavily on numerical methods.
These methods use various approaches to handle the immense Hilbert space of the environment, but generally struggle with simulating systems which also have big system Hilbert spaces. In this Paper, we present a method based on process tensors~\cite{jorgensen_Exploiting_2019} which greatly reduces the dependence of computational complexity on the system Hilbert space size, both time and memory-wise. This enables numerical simulation to tackle previously out-of-reach problems.

There already exists a rich landscape of numerical methods dedicated to studying non-Markovian open quantum systems~\cite{xu_Simulating_2026}. A few of these are equipped to handle larger systems, but they encounter other computational limitations. For example, wavefunction-based methods~\cite{MLMCTDH, chin_Exact_2010a} are generally better suited to larger system sizes since they only require acting on the Hilbert space of the system instead of the squared Liouville space. However they are limited by the poor quadratic scaling with number of time steps due to Lieb-Robinson type bounds~\cite{woods_Simulating_2015}. The recent  density matrix-based method Quantum accelerated stochastic propagator evaluation (Q-ASPEN)~\cite{grimm_Accurate_2024} was able to achieve better scaling with system size using gradient descent techniques to build tensor trains representing the environment degrees of freedom, but it was limited to short-lived correlations of the environment because of barren plateau effects~\cite{mcclean_Barren_2018}.

A compelling class of density matrix based methods are process tensor methods~\cite{Chiribella_2008,pollock_Non-Markovian_2018,keeling_Process_2025}, due to their wide range of application and their potential to unify other methods~\cite{ortega2024unifying,keeling_Process_2025}. 
The process tensor can be understood as the discretized Feynman-Vernon influence functional~\cite{feynman_theory_1963}, and crucially it can be efficiently compressed into a reusable matrix product operator (MPO) representation~\cite{jorgensen_Exploiting_2019}. 
Since the process tensor representing a given environment can be applied multiple times, such approaches allow for fast sweeps over system parameters~\cite{fux_Efficient_2021, mickiewicz2025exactfloquetdynamicsstrongly}, quick computation of multi-time correlation functions~\cite{dewit_Process_2025}, as well as efficient solving of optimal control problems~\cite{Butler2024, ortega2024unifying}.

Recently it has been shown that process tensor approaches can also radically improve the scaling of computational resources with number of simulation time-steps $n$, or with correlation lifetime $\tau$~\cite{fux_Process_2022, cygorek_Sublinear_2024, link_Open_2024}. Indeed, sublinear scaling in $n$ has been obtained with a $\order{\tau\log\tau}$ scaling in correlation lifetime by~\citet{cygorek_Sublinear_2024}. 
This has been further reduced to evaluating the process tensor in a cost that is independent of $n$  and scales  as $\order{\tau}$ by~\citet{link_Open_2024} using a time-translationally invariant process tensor (TTI-PT). This method has already proved useful to study quantum thermodynamics~\cite{shubrook_Numerically_2025}, driven-dissipative systems~\cite{kahlert_Simulating_2024} in cases where long-time dynamics are required, Floquet dynamics~\cite{mickiewicz2025exactfloquetdynamicsstrongly} and multi-time correlations~\cite{garbellini2026}.
While such methods dramatically reduce simulation time for small quantum systems, there remains a challenge of how computation time grows for large systems.

Some attempts to tackle large system sizes with process tensors have been made, but they require rather specific conditions. For example, when the system--environment coupling operator has multiple degenerate eigenvalues or degenerate pairwise eigenvalue differences~\cite{cygorek_Nonlinear_2017}, the effective system size can be reduced to the number of distinct eigenvalues. Other techniques include dealing with chains of open quantum systems with local dissipation~\cite{fux_Tensor_2023}, or using a mean-field treatment of the system in many-to-one coupling situations~\cite{fowler_Efficient_2022}.

In this paper we adapt the infinite time-evolving block decimation (iTEBD) algorithm~\cite{vidal_Classical_2007} used to build the TTI-PT in~\cite{link_Open_2024} to address the poor scaling with system Hilbert size.
The approach we describe is applicable regardless of the specific form of the coupling operator or of the system. For this purpose, we take advantage of the specific structure of the tensor network contracted through iTEBD, which features strongly compressible elements. This enables us to insert intermediate compression steps, resulting in greatly improved computation speeds and reduced memory use for the overall algorithm, with gains of orders of magnitude for systems over a dozen levels.

The rest of the paper is organized as follows:
In Sec.~\ref{sec:TTI}, we introduce the enhanced algorithm, and benchmark it against the prior version. In Sec.~\ref{sec:dispersive} we apply this new method to the dispersive readout problem in circuit QED, where a qubit coupled to a driven resonator leaks into a readout line which is treated as a non-Markovian environment, a problem previously beyond reach of non-Markovian methods.
Appendices provide further details of approximate analytic methods to which we compare our results. 

\section{Time-invariant process tensor approach}\label{sec:TTI}
In this section we introduce the improved iTEBD scheme.
This is based on the iTEBD algorithm introduced by~\citet{link_Open_2024}.
In order to make this discussion self contained, and to define the notation and diagrammatic approach needed, we begin in Sec.~\ref{sec:TTI:summary} by summarizing this previous work.
Section~\ref{sec:TTI:improved} then discusses the modifications to this algorithm which we introduce to improve scaling with Hilbert space dimension.

\subsection{Summary of iTEBD algorithm of Link, Tu, and Strunz}
\label{sec:TTI:summary}

We consider an open quantum system, where 
the system of interest governed by its Hamiltonian $\hat{H}_\mathrm{S}$ interacts with a Gaussian environment, which can be equivalently replaced by a bath of bosons interacting with the system. 
The full Hamiltonian for both the system and the environment is thus given by:
\begin{equation}\label{eq:OQS}
    \hat{H}_{SE} = \hat{H}_\mathrm{S}\otimes \mathds{1} + \hat{S}\otimes\sum_k(h_k^*\hat{b}_k +h_k\hat{b}_k^\dagger) + \mathds{1}\otimes\sum_k\omega_k\hat{b}_k^\dagger\hat{b}_k,
\end{equation}
where $\hat{S}$ is the system coupling operator, and $\hat{b}_k$ are the environment boson annihilation operators with corresponding frequency $\omega_k$ and coupling strength to the main system $h_k$. The environment is fully characterized by its spectral density $J(\omega)=\sum_k|h_k|^2\delta(\omega-\omega_k)$.
 
Since we are concerned with the dynamics of the system, we consider the evolution of the reduced system density matrix $\rho=\Tr_E(\rho_{SE})$ where $\rho_{SE}$ evolves under the Hamiltonian \eqref{eq:OQS} and $\Tr_E$ is the partial trace on the environment. For the rest of the paper, we consider the basis where $\hat{S}$ is diagonal and use Liouville-space indices. The density matrix is then represented by its elements $\rho_\mu=\rho_{\mu^l\mu^r}$ with $\mu = (\mu^l, \mu^r)$, where $\mu^l$ and $\mu^r$ index the system Hilbert space. We further use a Trotterized scheme to compute the time-evolution of this density matrix, which becomes exact in the limit of infinitesimally small time-step $\Delta t$.  The elements of the reduced density matrix after $n$ time-steps can be computed through:
\begin{equation}
    \rho_{\mu_n}(t_n) = \sum_{\mu_0,\dots,\mu_{n}}\mathcal{F}^{\mu_0,\dots,\mu_{n}}\prod_{i=1}^n \left(\mathcal{U}_{S,\mu_i\mu_{i-1}}\right)\rho_{\mu_0}(0)
\end{equation}
where $\mathcal{F}$ is the process tensor---a discretized version of the Feynman-Vernon influence functional~\cite{feynman_theory_1963}---and $\mathcal{U}_{S} = e^{-i\comm{\hat{H}_\mathrm{S}}{\cdot}\Delta t}$ is the system part of the evolution. In the case of a Gaussian environment with an initial state chosen as a product state between the bath in a thermal state and the system, the influence tensor $\mathcal{F}$ has a representation as a triangular tensor network~\cite{jorgensen_Exploiting_2019, strathearn_Efficient_2018a}:
\begin{figure}[t]
    \centering
    \includegraphics[width=\linewidth]{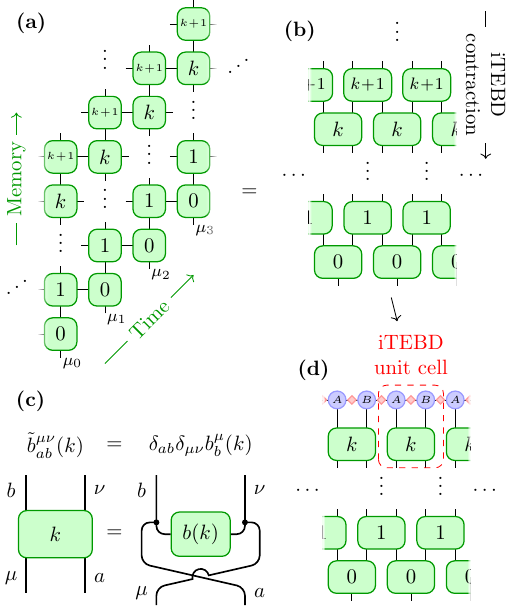}
    \caption{Graphical representation of the tensor networks for the time-invariant process tensors. (a) Two-dimensional infinite tensor network for the infinite influence tensor $\mathcal{F}^{\cdots\mu_0\mu_1\mu_2\mu_3\cdots}$ from~\cite{link_Open_2024}. (b) Equivalent rotated tensor network on which can be applied iTEBD from~\cite{link_Open_2024}. (c) Explicit representation of the $b^{\mu\nu}_{ab}(k)$ tensor with the Kronecker deltas as black dots. 
    (d) Tensor network illustrating the application of the iTEBD contraction scheme to the $\mathcal{F}$ tensor. The unit cell of the iTEBD step is highlighted by a red dashed rectangle. The step consists in applying the $\tilde{b}(k)$ gate to the infinite MPS in canonical form with its $A$, $B$ tensors in blue and the corresponding diagonal weight matrices as red diamonds.}
    \label{fig:scheme-TN}
\end{figure}
\begin{equation}\label{eq:triangle}
    \mathcal{F}^{\mu_0,\dots,\mu_{n-1}} = \prod_{i=0}^{n-1}\prod_{j=1}^i \left[b^{\mu_j}_{\mu_i}(i-j)\right],
\end{equation}
where $b$ is defined as:
\begin{equation}\label{eq:bks}
    \left[b^{\mu_j}_{\mu_i}(i-j)\right] = \exp\left[-(\lambda_{\mu_i^l}-\lambda_{\mu_i^r})(\eta_{i-j}\lambda_{\mu_j^l}-\eta_{i-j}^*\lambda_{\mu_j^r})\right],
\end{equation}
where $\lambda_{\mu}$ are eigenvalues of $\hat{S}$ and $(\mu_i^l, \mu_i^r) = \mu_i$ are the unraveled Hilbert space indices from the Liouville-space index $\mu_i$. The  coefficients $\eta_{i-j}$ are given by:
\begin{equation}
    \eta_{i-j}=\left\{\begin{aligned}\int_{t_{i-1}}^{t_i}\int_{t_{j-1}}^{t_j}\dd{t'}\dd{t''}C(t'-t''), \quad & i\neq j, \\ \int_{t_{i-1}}^{t_i}\int_{t_{i-1}}^{t'}\dd{t'}\dd{t''}C(t'-t''), \quad & i=j \end{aligned}\right.,
\end{equation}
where $C(t)$ is the environment correlation function:
\begin{equation}
C(t)=\int_0^\infty \dd{\omega}J(\omega)\left[\coth\!\left(\frac{\omega\beta}{2}\right)\cos(\omega t) - i \sin(\omega t)\right],
\end{equation}
with $\beta$ the inverse temperature.

The expression in Eq.~\eqref{eq:triangle} does not have an immediate representation as a two-dimensional tensor network but can be recast as one by expanding the $b$ tensors with additional legs~\cite{jorgensen_Exploiting_2019} by defining:
\begin{equation}
    \tilde{b}^{\mu\nu}_{ab}(k) = \left\{\begin{aligned}&\delta_{ab}\delta_{\mu\nu}b^\mu_b(k), &k > 0 \\ &\delta_{ab}\delta_{\mu\nu}\delta_{a\mu}b^\mu_b(0), &k=0\end{aligned}\right.,
\end{equation}
as represented in Fig.~\ref{fig:scheme-TN}c. Note that Greek indices correspond to legs that were vertical in the original orientation shown in Fig.~\ref{fig:scheme-TN}a, while Latin indices correspond to legs which were horizontal.
These four-legged tensors can then be contracted together and yield a triangular tensor network which is the starting point for various numerical methods~\cite{strathearn_Efficient_2018a, jorgensen_Exploiting_2019, link_Open_2024}. 
In particular, one may contract the tensor network to build a matrix product state (MPS) representation of the process tensor $\mathcal{F}$.

Importantly, recent work by~\citet{link_Open_2024} has shown it is possible to perform this contraction to find the  process tensor as a matrix product state consisting of one repeated time-independent tensor, yielding a TTI-PT. The contraction scheme used to obtain the TTI-PT consists in expanding the triangular network representing $\mathcal{F}$ to infinite times as shown Fig.~\ref{fig:scheme-TN}a. It can then be reshaped into a translationally-invariant two-dimensional network as in Fig.~\ref{fig:scheme-TN}b, that can be contracted using iTEBD~\cite{vidal_Classical_2007}.

As a quick summary, the iTEBD algorithm uses a two-site MPS which represents the unit cell of the infinite MPS. The MPS is represented in Fig.~\ref{fig:scheme-TN}d in the canonical gauge with the two site tensors $A$ and $B$ in blue and the diagonal weight matrices as red diamonds. At each step of the algorithm, gates---here our $\tilde{b}(k)$---are contracted with the two-site MPS. To prevent the exponential growth of the MPS bond dimension $\chi$, the MPS is truncated. It is then put back in the canonical gauge and the sites are swapped for the next iteration of the algorithm with the next gate $\tilde{b}(k-1)$. In order to apply iTEBD, the network is truncated at some memory depth $\tilde{k}$ by which point the bath correlation function $C(\tilde{k}\Delta t)$ should have decayed sufficiently. The algorithm then builds up the TTI-PT from this memory depth assuming an initial product MPS, until $\tilde{b}(0)$ is reached, giving the final process tensor. Contracting the network in this direction leaves the largest correlations $\eta_k$ till last, such that $\chi$ increases most at the end.

\subsection{Enhanced iTEBD scheme}
\label{sec:TTI:improved}

\begin{figure}[t]
    \centering
    \includegraphics[width=0.75\linewidth]{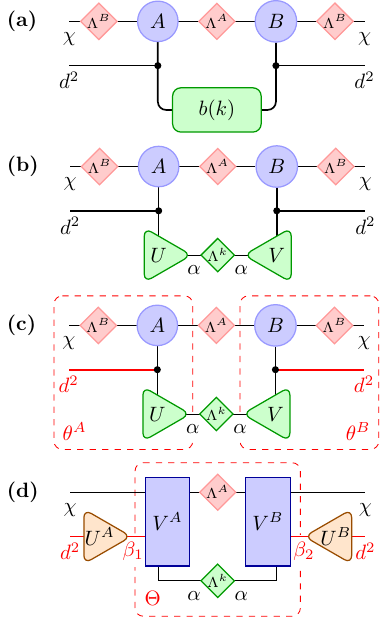}
    \caption{(a) Tensor network for the gate contraction of the enhanced iTEBD step. (b) Tensor network with SVD performed on the $b(k)$ tensor. (c) Tensor network displaying the partial SVD on the $\theta^A$ and $\theta^B$ blocks highlighted by red dashed rectangles, with respect to the red $d^2$ size leg. (d) Resulting tensor network from the partial SVDs. The final SVD is done on the $\Theta$ block highlighted by a red dashed rectangle.}
    \label{fig:scheme-iTEBD}
\end{figure}

The drawback of the iTEBD algorithm which we seek to address is the poor scaling with the Hilbert space size $d$ of the system. 
Each step of the iTEBD requires contracting the MPS and the gate together as shown in Fig.~\ref{fig:scheme-iTEBD}a, thus building a $\chi d^2\times\chi d^2$ matrix. This scales faster with $d$ than the memory requirement for the final process tensor which contains only $\chi^2 d^2$ elements. Moreover, it requires performing a singular value decomposition (SVD) on this matrix, in order to truncate the MPS, and to obtain the new $A$ and $B$ tensors. This SVD is time-consuming for large systems as its complexity grows as $\order{\chi^3 d^6}$. Since in practice $\chi$ will tend to grow with $d$ the scaling may be even worse than $d^6$. In this section we next present an improved scheme which avoids this bottleneck in the iTEBD contraction by taking advantage of the particular form of the $\tilde{b}$ gates and of the compressibility of $b$ tensors.

The $b(k)$ tensor is obtained by evaluating a function on the grid set by possible eigenvalue differences as per equation \eqref{eq:bks}. We may note that when the $\eta_k$ coefficients and the eigenvalue range of $\hat{S}$ are small enough, this function is an exponential of a low-rank polynomial with small values, which can itself can be approximated by a low-rank polynomial, making the $b(k)$ tensor very compressible~\cite{shi_Compressibility_2021}. Performing a first SVD on this matrix allows the truncation from $d^2$ elements to $\alpha$ which is often a fraction of $d^2$. The SVD on $b(k)$ drawn in Fig.~\ref{fig:scheme-iTEBD}b gives:
\begin{equation}
    b^\mu_b(k)=\sum_{i=1}^\alpha U_{bi}\Lambda^k_iV_{i\mu},
\end{equation}
where $U$ and $V$ are the resulting left and right unitaries, and $\Lambda^k$ is the diagonal matrix containing the relevant singular values.  We retain the label $k$ on $\Lambda$ to distinguish it from other matrices of singular values appearing in the calculation.

We then perform partial SVDs on the left and right parts of the network circled in red in Fig.~\ref{fig:scheme-iTEBD}c with respect to the $d^2$ size legs. For the left side we first define the relevant block as $\theta^A_{bilm}=\Lambda^B_{ll}A_{lbm}U_{bi}$, and the SVD is performed grouping the $ilm$ indices together. For the right side we have $\theta^B_{\mu ipm}=\Lambda^B_{pp}B_{mp\mu}V_{i\mu}$, and the SVD is performed grouping the $ipm$ indices together. After respective truncation to the first $\beta_1$ and $\beta_2$ singular values we get:
\begin{equation}
    \theta^A_{b[ilm]}=\sum_{q=1}^{\beta_1}U^A_{bq}V^A_{qilm}, \quad \theta^B_{\mu[ipm]}=\sum_{r=1}^{\beta_2}U^B_{\mu r}V^B_{ripm},
\end{equation}
where the singular values have been included into the right unitaries $V^A$ and $V^B$, and $U^A$ and $U^B$ are the respective corresponding left unitaries. The resulting network is drawn Fig.~\ref{fig:scheme-iTEBD}d.

Finally, the central iTEBD SVD is instead done on the contracted tensor network $\Theta$ which is circled in red Fig.~\ref{fig:scheme-iTEBD}d, excluding the unitaries $U^A$ and $U^B$:
\begin{equation}
    \Theta_{qrlp} = \sum_{i=1}^\alpha\sum_{m=1}^\chi V^A_{qilm}V^B_{ripm}\Lambda^A_{mm}\Lambda^k_{ii}.
\end{equation}
The SVD is thus performed on the reduced $\chi\beta_1\times\chi\beta_2$ $\Theta_{[qp][rl]}$ matrix instead of the full initial matrix. The leftover unitaries $U^A$ and $U^B$ are then contracted back only during the (c) step of the subsequent iteration, inside the $\theta$ blocks. This avoids ever building a tensor with $\chi^2d^4$ elements.

The key idea here---of performing partial SVDs to reduce the complexity of the large SVD---has recently also been discussed for the generic iTEBD algorithm, outside of the context of process tensors~\cite{xu_optimized_2023}. However, the specific form of the gates and the possibility of first reducing $b$ allows our algorithm to be even faster and more memory efficient: when $\alpha$ is much smaller than $d^2$, then the reduced inner bond dimensions $\beta_1$ and $\beta_2$ are also a fraction of $d^2$ as seen further in section~\ref{sec:benchmark}. We also investigated other choices of legs on which to perform the partial SVDs---notably since~\citet{xu_optimized_2023} performs the partial SVDs on the $\chi$-legs instead---but found our choice to be optimal in this case.

\subsection{Benchmarks}\label{sec:benchmark}
We now benchmark this method against the regular iTEBD algorithm for a system that consists of a harmonic oscillator, $\hat{H}_\mathrm{S}=\omega_r\hat{a}^\dagger\hat{a}$ with $\omega_r=1$, truncated to $d$ levels.  We choose the coupling operator to be a quadrature operator $\hat{S}=\hat{a}+\hat{a}^\dagger$. 
While we note that this model of non-interacting harmonic oscillators can in principle be exactly solved, we choose this problem for benchmarking for several reasons.
The model is close to that of interest in a variety of potential applications (such as that described in the next section), and the unequal spacing of eigenvalues of $\hat{S}$ makes it impossible to use previous techniques to build the process tensor more efficiently, such as the use of degeneracies described in Ref.~\cite{cygorek_Nonlinear_2017}.
We choose the frequently used ohmic spectral density with exponential cutoff defined as:
\begin{equation}
    J(\omega)=2\eta\omega\exp(-\omega/\omega_c),
\end{equation}
where $\eta$ is a dimensionless coupling strength and $\omega_c$ the frequency cutoff. Here, we take $\eta=0.01$, $\omega_c=3$, and $\Delta t = 0.1$. The parameters for the algorithm are the memory depth $\tilde{k}=500$ and $\epsilon_\mathrm{rel}=10^{-3}$ which selects that we keep the singular values with relative magnitude larger that $\epsilon_\mathrm{rel}$ as compared to the largest singular value in the last $\Theta$ SVD. This parameter controls the accuracy of the process tensor and has a large influence on the compute time and final bond dimension. The three other SVDs are truncated with relative accuracy $10^{-7}$ as will be the case throughout this paper.

\begin{figure}
    \centering
    \includegraphics[width=\linewidth]{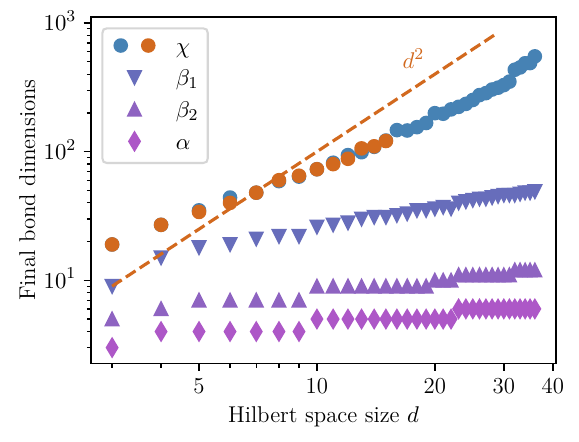}
    \caption{Bond dimensions for the last step of the TTI-PT computation. The final TTI-PT bond dimension $\chi$ is shown for the old method in orange and for the new method in blue. The intermediate bond dimensions of the new method $\alpha$, $\beta_1$ and $\beta_2$ are shown respectively with diamonds, down-facing triangles and up-facing triangles. These can be compared with the orange dashed $d^2$ line of the old method. The benchmarks are done for a harmonic oscillator coupled through the operator $\hat{a}+\hat{a}^\dagger$ to an ohmic bath with parameters $\eta=0.01$ and $\omega_c=3$ with iTEBD parameters $\tilde{k}=500$, $\epsilon_\mathrm{rel}=10^{-3}$.}
    \label{fig:bonds}
\end{figure}

Figure~\ref{fig:bonds} shows the bond dimensions for the last step of the algorithm run with these parameters as we vary $d$. The bond dimension in the last step is the most important for the computational cost because the bond dimensions are small during most of the iterations until the last few steps which are the most time and memory-consuming, just as in the regular algorithm~\cite{link_Open_2024}. The process tensor bond dimension $\chi$ for the regular and enhanced methods both agree, which is a good sign that no information is lost through the additional truncations of the enhanced algorithm. The figure clearly shows that  the intermediate bond dimensions $\alpha$, $\beta_1$, and $\beta_2$ are a lot smaller than the $d^2$ which would be reached without truncation. The potentially surprising difference between $\beta_1$ and $\beta_2$ is due to $b(k)$ not being symmetric, thus $U$ and $V$ have very different structures and information content.

\begin{figure}
    \centering
    \includegraphics[width=\linewidth]{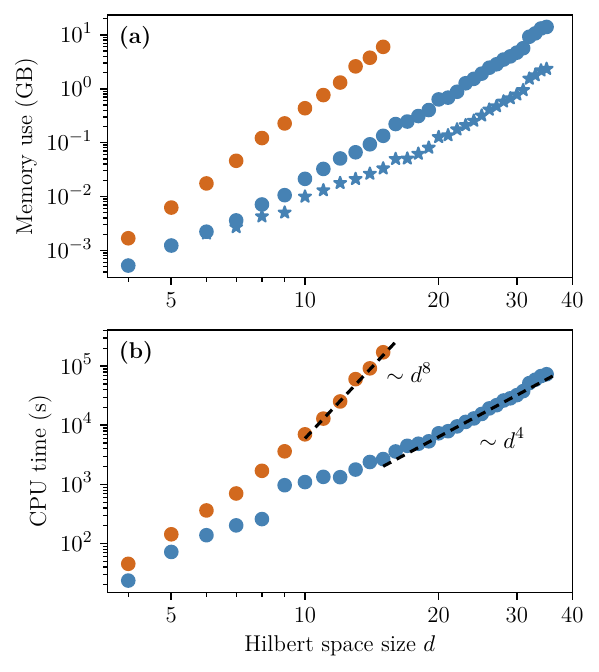}
    \caption{(a) Memory requirement to store the largest tensor as a function of Hilbert space size $d$. Comparison between the regular method (in orange circles) and the enhanced method (in blue circles). The stars represent the potential memory requirement with matrix-free methods for the $\theta$ SVDs. (b) Dependence of CPU computation time with Hilbert space size $d$ for the full algorithm. Comparison between the regular method (in orange) and the enhanced method (in blue). Approximate scaling for both methods are drawn in black dashed lines. The benchmarks are done for a harmonic oscillator coupled through the operator $\hat{a}+\hat{a}^\dagger$ to an ohmic bath with parameters $\eta=0.01$ and $\omega_c=3$ with iTEBD parameters $\tilde{k}=500$, $\epsilon_\mathrm{rel}=10^{-3}$. The computations were done using an Intel Xeon Gold 5218 Processor @ 3.9GHz with 32 cores.}
    \label{fig:benchmark}
\end{figure}

From these bond dimensions, we can determine the memory requirement for the generation of the TTI-PT. The largest tensors built during the algorithm are $\theta^A$ and $\theta^B$, which require $\chi^2d^2\alpha$ elements each, but do not have to be stored simultaneously. We plot the memory use of these tensors in Fig.~\ref{fig:benchmark}a, and compare them to the requirements for the $\chi^2d^4$ elements of the matrix of the original algorithm. We may note further that in the case where memory is the limiting factor, it is possible to use matrix-free methods to perform the truncated SVDs on the $\theta$ tensors without ever building them in memory. The largest tensor that must be stored during the algorithm would then be of size $\chi^2\max(d^2,\beta_1\beta_2)$ which is also displayed Fig.~\ref{fig:benchmark}a. For large $d$, it appears that $\beta_1\beta_2 < d^2$, and the memory requirement would then scale as $\chi^2d^2$ which has the same scaling as the final process tensor.

As for the time needed to run the full algorithm, the improvement is even more drastic, as seen in Fig.~\ref{fig:benchmark}b. The CPU time spent to generate the TTI-PT is lower than the original algorithm for all system sizes $d$ despite the overhead from the three additional SVDs. The scaling with all parameters is harder to estimate here due to all the various contractions and SVDs performed along the iTEBD, but we observe from the benchmarks an improvement from an approximate $\order{d^8}$ scaling to a $\order{d^4}$ scaling.

\section{Dispersive readout in circuit QED}
\label{sec:dispersive}
The improvement of scaling with Hilbert space size shown in the previous section opens up new possibilities in the types of problems that could be tackled in open quantum systems. 
To illustrate this, we consider a problem where the TTI-PT is particularly suitable: drive-induced dissipation in dispersive readout in circuit QED.

Dispersive readout of a superconducting qubit aims to read out the state of the qubit in a non-destructive way. 
It works by coupling a resonator to the qubit.
In the dispersive limit, the bare resonator frequency is shifted by an amount dependent on the qubit state. It is then possible to indirectly measure the qubit state by probing the resonator through a readout line~\cite{blais_Cavity_2004}.

A long-standing question in the field is to understand how driving the resonator impacts the readout, and especially the qubit lifetime $T_1$. Analytical derivations on simple models initially reported that increasing the drive strength should enhance $T_1$~\cite{boissonneault_Dispersive_2009, sete_Purcell_2014a}, whereas the experiments tend to show the opposite trend~\cite{minev_catch_2019}.

Although recent progress in the field suggests that this phenomenon could be due higher energy states of the full transmon Hamiltonian~\cite{hanai_Intrinsic_2021, dumas_MeasurementInduced_2024a}, we focus here on the qubit subspace, but consider the effects of a structured environment for decay from the resonator. We thus use a model that has very recently been developed by~\citet{Riva2026} that explicitly accounts for the deliberate modification of the transmission line spectral density using so-called Purcell filters.

Fully resolving the relative contributions of higher transmon levels vs structured transmission lines is beyond the scope of this paper, as simulations with the full transmon space require enhancements to the time propagation of process tensors to handle the very large system Liouville space of the combined transmon and resonator. This could be done for example by treating the coupled qubit-resonator system as a chain of two quantum systems could prove useful as efficient methods have recently been developed for propagating process tensors with chain systems~\cite{fux_Tensor_2023}.

Here, our motivation is to further demonstrate that our numerical approach now enables process tensor methods to treat driven-dissipative problems with large (composite) Hilbert spaces and structured spectral densities, which we have benchmarked using the detailed analysis of qubit--resonator--Purcell filter models introduced in~\cite{riva_Efficient_2026}.
In the following we first introduce the model we use to study this question in Sec.~\ref{sec:dispersive:model} and then present some results in Sec.~\ref{sec:dispersive:results}. 

\subsection{Dispersive readout model}
\label{sec:dispersive:model}

The diagram for the full circuit of our model is drawn Fig.~\ref{fig:scheme-circuit} and its Hamiltonian is given by:
\begin{equation}\label{eq:Hamiltonian}
    \hat{H} = \hat{H}_\mathrm{R} + \hat{H}_d + \hat{H}_\mathrm{E},
\end{equation}
which consists of the system Hamiltonian $\hat{H}_\mathrm{R}$ with a drive $\hat{H}_d(t)$, and the resonator environment Hamiltonian $\hat{H}_\mathrm{E}$. The circuit follows a common circuit QED setup~\cite{blais_Circuit_2021a} with a qubit at frequency $\omega_q$ transversely coupled to a resonator at frequency $\omega_r$ with coupling constant $g$. The Hamiltonian for this part of the system is the usual Rabi Hamiltonian given by:

\begin{figure}[t]
    \centering
    \includegraphics[width=\linewidth]{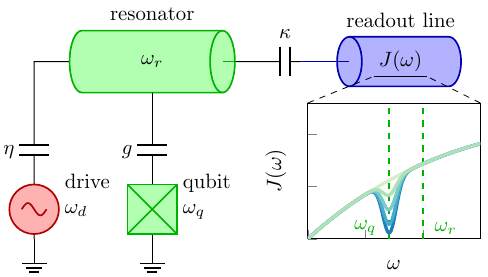}

    \caption{Circuit diagram for dispersive qubit readout representing the Hamiltonian \eqref{eq:Hamiltonian}. The Rabi Hamiltonian $\hat{H}_\mathrm{R}$ given by \eqref{eq:HR} is drawn in green, the drive Hamiltonian $\hat{H}_d$ given by \eqref{eq:Hd} is drawn in red, and the readout line Hamiltonian $\hat{H}_\mathrm{E}$ given by \eqref{eq:HE} is drawn in blue with the corresponding spectral densities pictured below in shades of green to blue for the Purcell filter strengths $p$ studied here: $0$, $0.25$, $0.5$, $0.75$, and $0.9$.}   
    \label{fig:scheme-circuit}
\end{figure}

\begin{equation}\label{eq:HR}
    \hat{H}_\mathrm{R} = \frac{\omega_q}{2}\hat{\sigma}_z + g \hat{\sigma}_x (\hat{a} + \hat{a}^\dagger) + \omega_r \hat{a}^\dagger\hat{a},
\end{equation}
where $\hat{\sigma}_x$ and $\hat{\sigma}_z$ are the qubit Pauli operators and $\hat{a}$ is the resonator annihilation operator. Most analytical derivations have been performed after making a rotating-wave approximation on this, leading to the Jaynes--Cummings Hamiltonian
\begin{equation}\label{eq:JC}
    \hat{H}_\mathrm{JC} = \frac{\omega_q}{2}\hat{\sigma}_z + g (\hat{\sigma}_+\hat{a} + \hat{\sigma}_-\hat{a}^\dagger) + \omega_r \hat{a}^\dagger\hat{a}
\end{equation}
instead. 
Recent derivations however show non-negligible differences between the rates produced by these two models~\cite{muller_Dissipative_2020}.
We discuss further below the differences that arise from these two models.

The resonator is driven through the exact non-RWA sine drive with strength $\epsilon$ and frequency $\omega_d$, whose Hamiltonian is:
\begin{equation}\label{eq:Hd}
    \hat{H}_d(t) = \epsilon\sin(\omega_d t)(\hat{a}+\hat{a}^\dagger).
\end{equation}
The resonator leaks into the readout transmission line which is capacitively coupled to the resonator and modeled as a Gaussian environment with a modified ohmic spectral density as per~\cite{yurke_InputOutput_2004, blais_Circuit_2021a}. The Hamiltonian of the readout line and its coupling to the resonator is thus given by:
\begin{equation}\label{eq:HE}
    \hat{H}_\mathrm{E} = (\hat{a} + \hat{a}^\dagger) \sum_k h_k (\hat{b}_k + \hat{b}_k^\dagger)+ \sum_k \omega_k\hat{b}_k^\dagger\hat{b}_k.
\end{equation}
We take the spectral density to have the form
\begin{equation}
    J_p(\omega)=\sum_k h_k^2\delta(\omega-\omega_k) = 2\eta\omega\exp(-\omega/\omega_c)H_p(\omega-\omega_q).
\end{equation}
In order to showcase the strength of this method and its capacity to go beyond simple smooth lineshapes, we will also model the Purcell filtering often used in circuit QED experiments by introducing an additional weight function $H_p(\omega)$ which we discuss below. If $H_p(\omega)=1$ then the spectral density is ohmic with strength $\eta$ and exponential cutoff with cutoff frequency $\omega_c$.

The Purcell effect, which is responsible for qubit relaxation induced by the resonator, is due to emission of resonator photons at the qubit frequency into the readout line~\cite{boissonneault_Dispersive_2009}. 
In that paper, an analytical perturbative expression for the qubit relaxation rate was derived for the Jaynes--Cummings model as:
\begin{equation}\label{eq:purcelljc}
    \gamma_\mathrm{JC}= \frac{1}{T_1} = \frac{g^2}{\Delta^2}J_p(\omega_q),
\end{equation}
equivalent to Fermi's golden rule.
Here $\Delta = \omega_q - \omega_r$ is the detuning; the factor $g/\Delta$ enters as describing the admixture of resonator into the qubit states in the dispersive limit.
Importantly the spectral density $J$ is evaluated at $\omega_q$ instead of the resonator frequency $\omega_r$ as previous derivations neglecting the frequency dependence of $J$ suggested~\cite{blais_Cavity_2004}. 
In order to enhance qubit $T_1$, some experiments apply a filter on the readout line at the qubit frequency to suppress dissipation by reducing $J_p(\omega_q)$. The effective relaxation rate of the qubit is thus reduced without impacting the resonator. The two most frequently used filters are bandpass filters at the resonator frequency~\cite{jeffrey_Fast_2014} or notch (band-rejection) filters at the qubit frequency~\cite{reed_Fast_2010}. Here we will focus on the latter, which we model with the notch function $H_p(\omega) = 1-p\exp(-\omega^2/w^2)$ with $1/w^2=150$ where $p$ gives the strength of the filter: $J_p(\omega_q)=(1-p)J_{p=0}(\omega_q)$. The filtered spectral densities are shown in the inset of Fig.~\ref{fig:scheme-circuit} for the various values of $p$ considered here.

\subsection{Numerical results}
\label{sec:dispersive:results}

We now turn to presenting numerical results on simulating the full Rabi model given in Eq.~\eqref{eq:Hamiltonian}.
The first step for the numerical simulations is the generation of the TTI-PT for the readout line environment $\hat{H}_\mathrm{E}$. 
Because the system includes both the qubit and the resonator, the size of the system Hilbert space is $2N$ where $N$ is the truncation level of the resonator space. Since the coupling operator $(\hat{a}+\hat{a}^\dagger)$ only acts on the resonator subspace, it is however possible to generate the TTI-PT for only the $N$-level resonator. During the Trotterized time-evolution, the density matrix is represented as $\rho_{\mu_n\nu_n}$, where $\mu_n$ and $\nu_n$ are the respective Liouville-space indices for the combined resonator and qubit. The TTI-PT is then only applied to the resonator leg of the density matrix---such an approach is analogous to the original use of degeneracies to simplify path summation~\cite{cygorek_Nonlinear_2017}.
In appendix~\ref{appendixA} we check that the process tensors generated with this new method give rise to the right dynamics when applied to an exactly solvable model, excluding the qubit-resonator coupling.

\subsubsection{Purcell decay rates}
\label{sec:numres:purcell}

\begin{figure}[t]
    \centering
    \includegraphics[width=\linewidth]{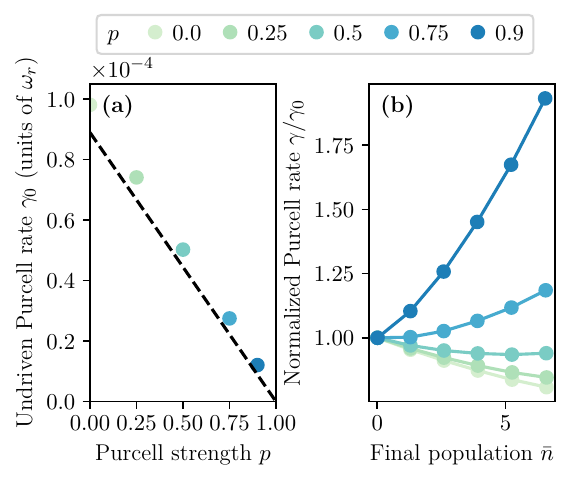}
    \caption{(a) Qubit relaxation rates at zero drive $\gamma_0$ as a function of filter strength $p$ as dots compared to the theoretical value from Eq.~\eqref{eq:purcell} as a black dashed line. (b) Normalized qubit relaxation rates $\gamma/\gamma_0$ as a function of the final average population in the resonator for multiple values of the Purcell filter strength $p$ drawn in shades of green to blue ($p=0$, 0.25, 0.5, 0.75, 0.9). Simulations were run with the full Hamiltonian \eqref{eq:Hamiltonian} and TTI-PTs generated using $\tilde{k}=1000$, $\epsilon_\mathrm{rel}=10^{-7}$, $\Delta t=2\pi\omega_r/62$, and a truncation to the $N=20$ first resonator states. 
    Simulations were run for times up to $t=5(\eta\omega_r)^{-1}$ and Purcell rates where extracted from fitting the dynamics of $\langle\hat{\sigma}_z(t)\rangle$ to a decaying exponential over the last third of the simulation. System parameters are experimentally relevant values in the strong coupling regime: $g=\SI{0.211}{\giga\hertz}$, $\omega_q=\SI{5.304}{\giga\hertz}$, $\omega_r=\SI{7.5}{\giga\hertz}$. The spectral density has parameters $\eta=10^{-3}$, $\omega_c=3\omega_r$, giving a loss-rate for the resonator of around $\kappa=\SI{0.068}{\per\nano\second}$.}
    \label{fig:purcell-rates}
\end{figure}

Figure~\ref{fig:purcell-rates} shows the behavior of the effective decay rate found from these calculations.
Figure~\ref{fig:purcell-rates}a first shows the extracted rates $\gamma_0$ for the different Purcell filter strengths $p$ in the undriven case. 
As expected, the Purcell filters lower the relaxation rate $\gamma_0$.
The drive-dependent behavior of the relaxation rates is then shown in Fig.~\ref{fig:purcell-rates}b for various Purcell filters. We find that for the unfiltered ohmic spectral density, the relaxation rate decreases when drive power is increased, as expected for a two-level truncated transmon. For stronger filters, the behavior is reversed such that increasing the drive also increases dissipation, which is closer to experimental trends.

The likely physical intuition behind this phenomenon is that this is due to ac-Stark shift. When driving the resonator, the qubit frequency is detuned from its bare frequency $\tilde{\omega}_q=\omega_q+\chi(2\bar{n}+1)$ where $\chi=g^2/\Delta^2$ is the dispersive shift~\cite{blais_Circuit_2021a}. The Purcell relaxation of the qubit would thus happen by releasing resonator photons at frequency $\tilde{\omega}_q$, and the Purcell rate would be proportional to $J_p(\tilde{\omega}_q)$ instead. The resulting analytical rates are superimposed in appendix \ref{appendixC} and, although not a perfect match, suggest that the explanation captures the right physical effect. The remaining differences can be attributed to non-Markovian effects, or to further effects of the drive which not only shifts the qubit frequency, but also broadens the qubit linewidth.

The idea that shift and broadening effects could be an explanation for drive-induced qubit lifetime suppression has also been discussed in Ref.~\cite{thorbeck_ReadoutInduced_2024}.
However, that paper proposes that the effect is rather due to high spots in the spectral density of the qubit environment instead of effects from the resonator environment,  as discussed in our work.

To gain further understanding into the behavior discussed above, we can also compare the process tensor results to approximate Markovian simulations.
This is shown in appendix~\ref{appendixB} for the assumptions of a flat and weak spectral density and validity of the Jaynes--Cummings approximation. 
As discussed in that appendix, under those conditions, there is a simple Lindblad equation with only resonator photon loss.
However, this simple Lindblad equation stops working as soon as one of the assumptions above is lifted. 
In particular, the qubit relaxation rate predicted by the Lindblad equation is not sensitive to whether Rabi or Jaynes--Cummings Hamiltonian are being used~\cite{Gogoi2026}, nor to any deviation of the spectral density near the qubit frequency compared to the value at the resonator frequency.

Within the time-local Markovian approach, one can derive an improved master equation by making use of the full system Hamiltonian to define eigenoperators of the system.
This was the approach used in Ref.~\cite{boissonneault_Dispersive_2009} to derive \eqref{eq:purcelljc}.
More recently a similar approach has been formulated using a diagrammatic expansion as in~\cite{muller_Dissipative_2020},
which obtains a correction to the Purcell rate formula for the Rabi Hamiltonian:
\begin{equation}\label{eq:purcell}
    \gamma = \left(\frac{2\omega_r}{\omega_r+\omega_q}\right)^2\gamma_\mathrm{JC} = \frac{4g^2\omega_r^2}{(\omega_q^2-\omega_r^2)^2}J_p(\omega_q).
\end{equation}
Unfortunately, these expansions become challenging when taking into account the full system Hamiltonian including the drive, and their validity domain is confined to small expansion parameters. Moreover, when introducing sharp features like a notch, it is necessary to be careful as Lindblad dissipator expansions are likely to fail to capture the correct dynamics~\cite{stefanini_Lindblad_2025}. This is where exact numerical methods come into play.

We compare results expected from this corrected Purcell rate to both undriven rates as shown Fig.~\ref{fig:purcell-rates}a by the dashed line, and to driven rates in Fig.~\ref{fig:stark-shift} of appendix~\ref{appendixC} with dashed lines as well. In both cases, there is indeed qualitative agreement with our numerical simulations, but obtaining exact rates requires the full non-Markovian simulations.

\subsubsection{Resonator state}
\label{sec:numres:prob}
\begin{figure}[b]
    \centering
    \includegraphics[width=\linewidth]{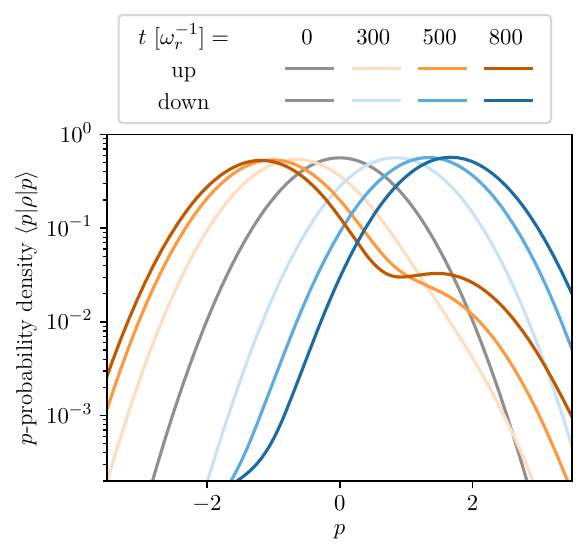}
    \caption{$p$-probability distributions $\bra{p}\rho\ket{p}$ of the resonator at various times ($\omega_r^{-1}t=0$, $300$, $500$, $800$) of the dynamics, starting with a resonator vacuum and either up or down qubit. $\hat{p}=i(\hat{a}-\hat{a}^\dag)$ is a quadrature operator with eigenstates $\ket{p}$. Distributions from runs with initial states with up qubits are drawn in orange and ones with down qubits are drawn in blue. The initial vacuum distribution of time zero is drawn in gray. The model used here is the full Hamiltonian \eqref{eq:Hamiltonian} with ohmic spectral density (no Purcell filter) and identical parameters to Fig.~\ref{fig:purcell-rates}.}
    \label{fig:p-distribution}
\end{figure}

We can also note that since our approach allows direct access to the resonator density matrix, it is possible to obtain experimentally interesting observables like readout fidelity. Experimentally this is computed by compiling the single-shot quadrature readouts into a histogram depending on the initial qubit state~\cite{walter_Rapid_2017}. The crossing between the histograms for initial up and down states gives the decision boundary for future experiments for determining the qubit state from the single-shot quadrature. The readout infidelity or error is given by the overlap between the two histograms. These histograms are readily available from our simulations as seen in Fig.~\ref{fig:p-distribution}.

As time increases, the separation between the histograms increases giving better readout fidelity, until at late times, the qubit decay due to the Purcell effect becomes too large. The initial up-state histogram in orange indeed shows that for later times, the distribution is contaminated by switching of the qubit to the down state as per Purcell relaxation, characterized by a second peak appearing where the down-state histogram peaks.

\section{Conclusion}
In this work, we have obtained a new method for drastically reducing the computational cost of building process tensors for large systems, going beyond the currently existing methods which are restricted to coupling operators with degenerate eigenvalues~\cite{cygorek_Nonlinear_2017} or chain-like systems~\cite{fux_Tensor_2023}. The computation time scaling of process tensor methods with system size was brought down from $\order{d^8}$ as in both the original PT-MPO algorithm by~\citet{jorgensen_Exploiting_2019} and the original TTI-PT algorithm by~\citet{link_Open_2024} to approximately $\order{d^4}$. We achieved this significant improvement by exploiting the compressible structure of the blocks in the influence tensor network.

The need for such optimization had been identified in the literature~\cite{cygorek_ACE_2024}, drawing motivation from the Q-ASPEN method~\cite{grimm_Accurate_2024} which used Chebyshev interpolation to construct tensor trains decompositions for their \textit{influence kernel}, allowing for simulations of much larger system sizes. Here, we instead used a regular SVD to compress the blocks, and follow an iTEBD algorithm for the full compression, which has the advantage of not requiring training of the decomposition through gradient descent. The use of gradient descent, and associated barren-plateau problem~\cite{mcclean_Barren_2018} limited the Q-ASPEN algorithm to a few steps of memory depth while our method is capable of thousands of time steps of memory depth.

We applied this new method to the problem of Purcell decay in circuit QED which was unreachable with previous process tensor based methods, and which proved difficult for methods which are better optimized for larger systems such as the Time Evolving Density operator with Orthogonal Polynomials Algorithm~\cite{chin_Exact_2010a} due to the long simulation times required. We showed that the TTI-PTs generated through our optimized method were capable of accurate simulations through tens of thousands of time steps of large driven systems of more than 30 levels. Moreover, the system coupling operator used here was the quadrature $\hat{a}+\hat{a}^\dagger$ operator, which is ubiquitous in quantum systems, and whose non-degenerate spectrum did not permit previous degeneracy-based approaches. Coupling through such an operator is also very common in open quantum systems as it is a result of the reaction coordinate (RC) mapping~\cite{Nazir2018}. This mapping brings any coupling to a Gaussian environment back to coupling to a bosonic system, which is itself coupled to a bath through this operator---with a coupling strength independent of the initial coupling between the system and its environment. This opens prospects for simulations of strongly coupled systems, as well as structured spectral densities with a dominant mode which would otherwise incur long-lived time correlations which are known to be hard to deal with for process tensors~\cite{cygorek_ACE_2024}. One could also investigate time-dependent couplings to a bath as the RC would bring this time-dependence back into the system.

Finally, the approach we present in this paper gives the possibility of using all the benefits of process tensors, and more specifically TTI-PTs, to study larger systems: from computation of multitime correlation functions~\cite{dewit_Process_2025,garbellini2026} to optimization problems---be they over a set of parameters~\cite{mickiewicz2025exactfloquetdynamicsstrongly} or full optimal control~\cite{fux_Efficient_2021}---through simple time evolution as in the present work. Further work is needed to understand whether such optimizations are also possible for the regular PT-MPO algorithm, or for the ACE algorithm. Moreover, diagonalizing or even building the combined system--TTI-PT propagator---obtained by contracting the TTI-PT and system propagators---which can be used to speed up computations with TTI-PTs specifically \cite{link_Open_2024, mickiewicz2025exactfloquetdynamicsstrongly}, can become untractable for large system sizes. Additional techniques might thus be required to fully take advantage of the benefits of TTI-PTs.

\section{Data Availability}

All simulations where ran using the open source package \textsc{OQuPy}~\cite{OQuPy, fux_tempoCollaboration_2024}.
\begin{acknowledgments}
We acknowledge helpful discussions with Alex Petrescu, Angela Riva, Prakritish Gogoi and Roosmarijn de Wit.
E.C. acknowledges funding from ENS de Lyon.
A.W.C.  wishes to acknowledge support from the ANR Project RADPOLIMER (Grant No. ANR-22-CE30-0033)
This work used the Cirrus UK National Tier-2 HPC Service at EPCC (http://www.cirrus.ac.uk) funded by The University of Edinburgh, the Edinburgh and South East Scotland City Region Deal, and UKRI via EPSRC.
A.W.C, J.K., and B.W.L. acknowledge support from the Seed Meeting program of the Higher Education, Research and Innovation Department of the French Embassy in the UK.

\end{acknowledgments}

\appendix
\section{Comparison of TTI-PT to exact dynamics}\label{appendixA}

In this appendix we demonstrate the validity of the TTI-PTs produced with our method by benchmarking the dynamics they produce against exact methods. For the driven-dissipative resonator without the transmon, the Hamiltonian is quadratic and thus exactly solvable. The coefficients of this quadratic Hamiltonian are obtained by chain-mapping of the continuum of bath modes into a discrete chain of modes which can be truncated to obtain exact dynamics until a set time (proportional to the number of modes kept)~\cite{chin_Exact_2010a, lacroix_MPSDynamicsjl_2024a}. The coefficients allow one to build the corresponding Bogoliubov Hamiltonian which is then diagonalized to obtain time-evolved expectation values of operators of interest. Here we focus on two important operators: the quadrature operator $\hat{a}+\hat{a}^\dagger$ and the population operator $\hat{n}=\hat{a}^\dagger\hat{a}$ of the resonator.

\begin{figure}[b]
    \centering
    \includegraphics[width=\linewidth]{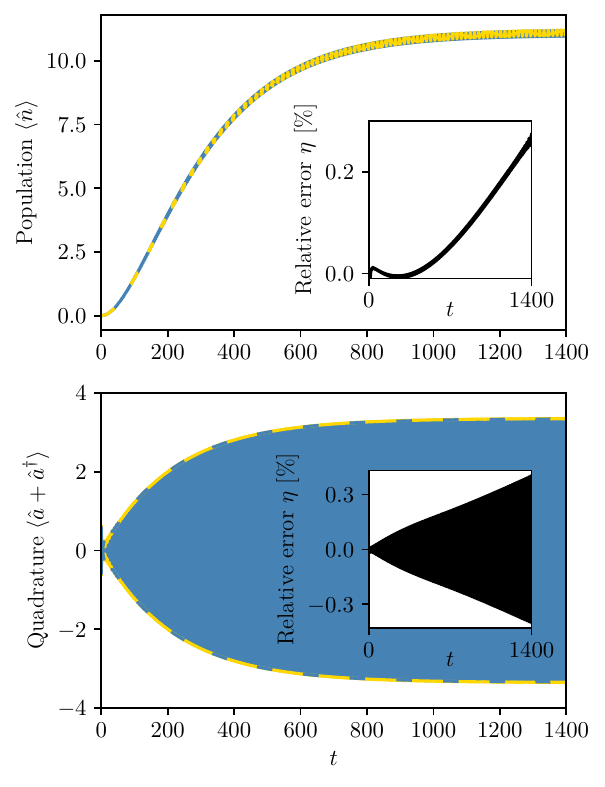}
    \caption{Dynamics in the driven-dissipative resonator model with ohmic spectral density. (a) Population dynamics $\langle\hat{n}(t)\rangle$ for TTI-PT in blue compared to the exact dynamics in gold dashes. Inset shows relative error between process tensor simulations and exact dynamics. (b) Quadrature operator dynamics $\langle(\hat{a}+\hat{a}^\dagger)(t)\rangle$ for TTI-PT in blue. Envelope of the exact dynamics is shown in gold dashes. Inset shows relative error between process tensor simulations and exact dynamics.}
    \label{fig:bare_cavity}
\end{figure}

The system Hamiltonian is taken to be the full Hamiltonian \eqref{eq:Hamiltonian} without the qubit:
\begin{equation}
    \hat{H}_\mathrm{S}(t) = \omega_r \hat{a}^\dagger\hat{a} + \epsilon\sin(\omega_dt)(\hat{a}+\hat{a}^\dagger),
\end{equation}
with $\omega_d$ chosen such that the driving is on resonance. The frequency of the resonator is extracted from exact dynamics of the undriven resonator with its environment by fitting the dynamics to a decaying sinusoid. $\eta=0.03\omega_r$ is chosen such that the resonator has an average of at least 10 photons in the steady state. The environment spectral density is ohmic (without filter) with exponentially decaying cutoff with the same parameters as in the main text: $\eta=0.001$, $\omega_c=3\omega_r$.

The TTI-PT for this benchmark is generated with $\epsilon_\mathrm{rel}=10^{-7}$ and $\tilde{k}=1000$, $\Delta t=2\pi/62\approx 0.1$ in units where $\omega_r = 1$ (these are the same parameters as used throughout this paper). Here we take $d=30$ for the truncation of the resonator space. The simulations were run until the resonator reaches a driven-dissipative steady state which takes time of order $(\eta\omega_r)^{-1}$. 
For this comparison, 4200 modes were used in the exact method (requiring diagonalization of a 8400$\times$8400 matrix).

The results for both populations and the quadrature operator seem to match perfectly according to Fig.~\ref{fig:bare_cavity}, although the oscillations are not resolved since they are too fast compared to the total simulation time. The insets reveal that across these simulations, the relative error is around $0.3\%$ and increases with time. Extrapolating to the simulation times used to extract the Purcell rates gives less than a percent error on the dynamics.

\section{Comparing drive-induced decay to Stark-shifted}\label{appendixC}
\begin{figure}[h]
    \centering
    \includegraphics[width=\linewidth]{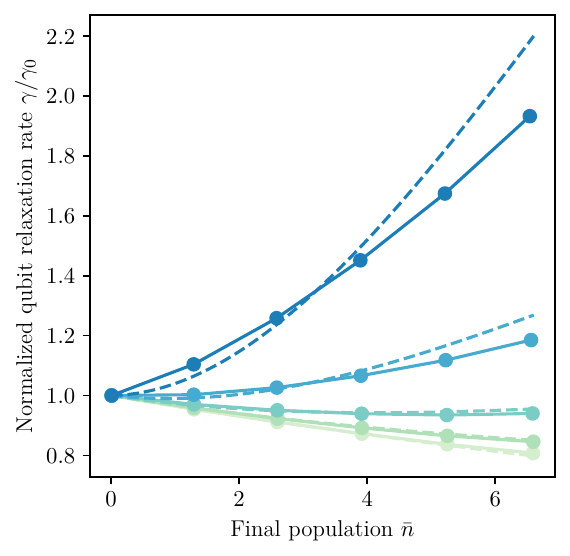}
    \caption{Normalized qubit relaxation rates $\gamma/\gamma_0$ as a function of the final average population in the resonator for multiple values of the Purcell filter strength $p$ drawn in shades of green to blue ($p=0$, 0.25, 0.5, 0.75, 0.9). The exact values from the simulations are drawn with full lines and the approximate values from the Stark-shift formula are drawn in dashed lines.}
    \label{fig:stark-shift}
\end{figure}

Here, we compare the normalized Purcell rates from the simulations to what could be expected due to evaluating the Markovian expression for qubit decay accounting for Stark-shifting of the qubit frequency:
\begin{equation}
    \frac{\gamma}{\gamma_0} = \frac{(\omega_0^2-\omega_r^2)^2 J_p(\tilde{\omega}_q)}{(\tilde{\omega}_q^2-\omega_r^2)^2 J_p(\omega_0)},
\end{equation}
where $\tilde{\omega}_q=\omega_q+\chi(2\bar{n}+1)$ is the Stark-shifted frequency, $\bar{n}$ denotes the average photons in the resonator and $\omega_0 =\omega_q+\chi$ is the 0-drive Stark-shifted frequency.
The simulation results agree reasonably well this intuitive formula for small filter depth, but start moving away for deeper filters---or sharper spectral densities---as shown Fig.~\ref{fig:stark-shift}. This is expected since this intuitive formula uses \eqref{eq:purcell} which is derived from a Lindblad expansion~\cite{muller_Dissipative_2020}. It is however interesting to note that this formula seems to predict the accurate behavior for the smoothest spectral densities whereas previous studies on the Lindblad model \eqref{eq:lindblad} showed that this kind of Stark-shift formula did not match the simulations nor the theory quantitatively~\cite{sete_Purcell_2014a}.

\section{Recovering Lindblad dynamics for a flat spectral density}\label{appendixB}
\begin{figure}[h]
    \centering
    \includegraphics[width=\linewidth]{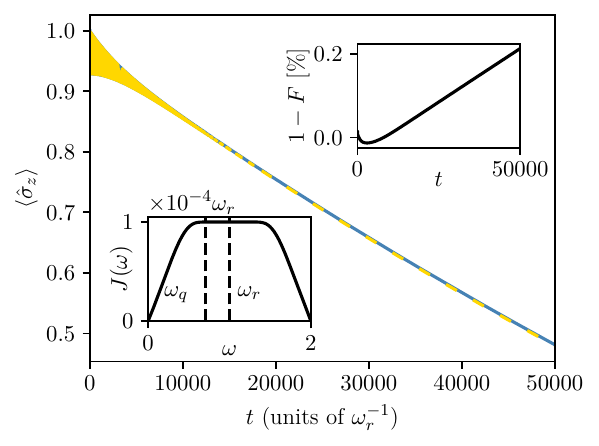}
    \caption{Qubit dynamics $\langle\hat{\sigma}_z(t)\rangle$ in the Jaynes--Cummings model with flat spectral density. The TTI-PT results are drawn in blue compared to the Lindblad dynamics in gold dashes. One inset shows the error $1-F$ between process tensor and Lindblad simulations where $F$ is the quantum state fidelity between the density matrices obtained from both simulations. Other inset shows the flat spectral density $J(\omega)$ (in $\omega_r$ units) used for the TTI-PT simulation. Parameters used for the simulation are the spectral density of Eq.~\eqref{eq:flat-sd} with $\eta=10^{-4}$, identical Jaynes--Cummings Hamiltonian parameters as in the main text, and parameters used to generate the TTI-PT are $\tilde{k}=1000$ and $\epsilon_\mathrm{rel}=10^{-7}$ with truncation of Hilbert space to $N=13$ resonator states.}
    \label{fig:flat-sd}
\end{figure}
Many previous numerical studies of this problem used a Lindblad approach~\cite{sete_Purcell_2014a}. This approach uses the Lindblad model for the resonator while keeping the same system Hamiltonian, giving the following evolution operator for the density matrix:
\begin{gather}\label{eq:lindblad}
    \mathcal{L}\rho = -i[\hat{H}_\mathrm{JC}, \rho] + \kappa\mathcal{D}[\hat{a}](\rho), \\ \mathcal{D}[\hat{a}](\rho) = \hat{a}\rho\hat{a}^\dagger-\frac{1}{2}(\hat{a}^\dagger\hat{a}\rho+\rho\hat{a}^\dagger\hat{a}),
\end{gather}
where $\mathcal{D}[\hat{a}]$ is the resonator loss dissipator with loss rate $\kappa=2\pi J(\omega_r)$. The assumptions required to obtain this equation are the following: the spectral density should be small enough that the Markov approximation can be made, the Jaynes--Cummings approximation for the Hamiltonian should be used (i.e.\ the RWA approximation should be valid), and the spectral density should be flat so that the qubit can be added without probing rates other than $\kappa$. If one of these approximations is not valid, it is then instead necessary to derive the Lindblad equation from the full system Hamiltonian which is more challenging than just adding a single dissipator for the resonator.

We show, using a small and nearly flat spectral density, that the Lindblad and TTI-PT simulations agree when these hypotheses are met. The chosen spectral density is:
\begin{equation}\label{eq:flat-sd}
    J(\omega) = \eta\omega_r \left(1-\exp(1-\frac{\omega_r^2}{(\omega-\omega_r)^2})\right), \omega\in[0, 2\omega_r],
\end{equation}
where $\eta=10^{-4}$, which is almost flat near $\omega_r$ with constant value $\eta$. With qubit and resonator frequency values taken as in the main text, $J$ is sufficiently flat so that $J(\omega_q)$ is in within $0.001\%$ of $J(\omega_r)$.

The resulting dynamics obtained with a TTI-PT almost perfectly match the expected Lindblad evolution. The comparison between qubit dynamics is shown in Fig.~\ref{fig:flat-sd}, where the curves overlay nearly perfectly. The infidelity between the two density matrices is less that $0.2\%$ over this simulation, showing that the TTI-PT simulation does recover the Lindbad approximation when the right hypotheses are met. The extracted Purcell rate from fitting these dynamics is within $1\%$ of the theoretical rate computed with \eqref{eq:purcelljc}.

\bibliography{biblio}
\end{document}